\documentclass[aps,onecolumn,notitlepage,groupedaddress,superscriptaddress,amsmath,amssymb,showpacs,noeprint, PRA]{revtex4-2}

\usepackage{graphicx}
\usepackage{dcolumn}
\usepackage{bm}
\usepackage{hyperref}
\usepackage{mathtools, cuted}
\usepackage{verbatim}
\usepackage[mathlines]{lineno}
\usepackage{physics}
\usepackage{xfrac}

\usepackage{soul}
\usepackage{color}
\usepackage{csquotes}
\usepackage{comment}

\newcommand{\prt}[1]{\left(#1\right)}

\newcommand{\prtg}[1]{\left\{#1\right\}}

\begin{document}
	
\title{Extraction of Work via a Thermalization Protocol \footnote[1]{Presented at the 11th Italian Quantum Information Science conference (IQIS2018), Catania, Italy,
17-20 September 2018.}}

\author{Nicol\`o Piccione}
\affiliation{Institut UTINAM - UMR 6213, CNRS, Universit\'{e} Bourgogne Franche-Comt\'{e}, Observatoire des Sciences de l'Univers THETA, 41 bis avenue de l'Observatoire, F-25010 Besan\c{c}on, France}

\author{Benedetto Militello}
\affiliation{Dipartimento di Fisica e Chimica - Emilio Segr\`{e}, Universit\`a degli Studi di Palermo, via Archirafi 36, I-90123 Palermo, Italy}
\affiliation{INFN Sezione di Catania, via Santa Sofia 64, I-95123 Catania, Italy}

\author{Anna Napoli}
\affiliation{Dipartimento di Fisica e Chimica - Emilio Segr\`{e}, Universit\`a degli Studi di Palermo, via Archirafi 36, I-90123 Palermo, Italy}
\affiliation{INFN Sezione di Catania, via Santa Sofia 64, I-95123 Catania, Italy}

\author{Bruno Bellomo}
\email{bruno.bellomo@univ-fcomte.fr}
\affiliation{Institut UTINAM - UMR 6213, CNRS, Universit\'{e} Bourgogne Franche-Comt\'{e}, Observatoire des Sciences de l'Univers THETA, 41 bis avenue de l'Observatoire, F-25010 Besan\c{c}on, France}

\begin{abstract}
This extended abstract contains an outline of the work reported at the conference IQIS2018. We show that it is possible to exploit a thermalization process to extract work from a resource system $R$ to a bipartite system $S$. To do this, we propose a simple protocol in a general setting in the presence of a single bath at temperature $T$ and then examine it when $S$ is described by the quantum Rabi model at $T=0$. We find the theoretical bounds of the protocol in the general case and we show that when applied to the Rabi model it gives rise to a satisfactory extraction of work and efficiency.
\end{abstract}

\maketitle

\section{Introduction}

In this extended abstract, based on {Ref.}\cite{Piccione(2018)}, we present a work-extraction protocol which exploits a simple thermalization process as the main ingredient.
To this purpose, it is necessary to define the quantity work and the meaning of ``work extraction''.
Defining work is a difficult task in quantum mechanics and a general definition has not yet been adopted~\cite{Perarnau-Llobet(2017)}.
We base our protocol on a definition~\cite{Gallego2016} given inside the theoretical framework of thermodynamic resource theory. This deals with operations performed on a system and a thermal bath, by imposing that the sum of the energies of the system and the bath is conserved through the process (not only on average but also in the entire distribution) \cite{Horodecki2013}. \mbox{This recently} developed theoretical framework provided interesting results such as generalizations of the second law of thermodynamics~\cite{Brandao(2015),Lostaglio(2015)} and a way to exploit  coherences in the energy basis~\cite{Aberg2014}.

In the definition of Ref.~\cite{Gallego2016}, we need to fix the temperature of the thermal bath and divide the system of interest in two systems.
We name one of them $R$, the resource from which we want to extract the work, and the other one $S$, the system on which we want to do work.
The work done or extracted is quantified by the difference of a quantity, which is inherent to the state of the system $S$, between the start and the end of the process.
The work quantifier can be chosen among different quantities and its choice depends on the physical constraints that we consider for the process at hand.
Furthermore, whatever the chosen quantity is, the amount of work extracted from system $S$ has to be always equal or lower than the amount of work lost by system $R$.
In Ref.~\cite{Gallego2016} there are a lot of possible quantifiers of work done on (or extracted from) a system, among which we choose to adopt the following one:
\begin{equation}
\label{eq: Eisert definition}
W=\Delta F\prt{\rho',H'}-\Delta F \prt{\rho,H},
\end{equation}
\noindent where
\begin{equation}
\label{eq: definition delta F}
\Delta F\prt{\rho,H}=F\prt{\rho,H}-F\prt{\rho^{\textup{th}},H } ,\quad F\prt{\rho,H}=\Tr \prt{H\rho} -k_B TS\prt{\rho}\quad \mathrm{and} \quad
S(\rho) =-\Tr \prtg{\rho \ \ln \prt{\rho}}.
\end{equation}

Here, $F\prt{\rho,H}$ is the free energy of the state $\rho$ when the system is governed by the Hamiltonian $H$, $\rho^{\textup{th}}$ is the thermal state of the system at temperature $T$ equal to the temperature of the thermal bath which is used in the process, $k_B$ is the Boltzmann's constant and $S\prt{\rho}$ is the Von Neumann entropy of the state $\rho$.
The quantities which have the symbol $'$ are connected to the end of the process, while those not marked are connected to the start of the process.
If $H'=H$, Equation~\eqref{eq: Eisert definition} reduces to the \mbox{simpler form}
\begin{equation}
\label{eq: work definition no Hamiltonian change}
W=F\prt{\rho',H}-F\prt{\rho,H}.
\end{equation}

Contrarily to other definitions of work, this one is not based on measurements.

\section{Results and Discussion}

\subsection{The Thermalization Protocol}

\begin{figure}[t!]
	\centering
	\includegraphics[width=0.5\textwidth]{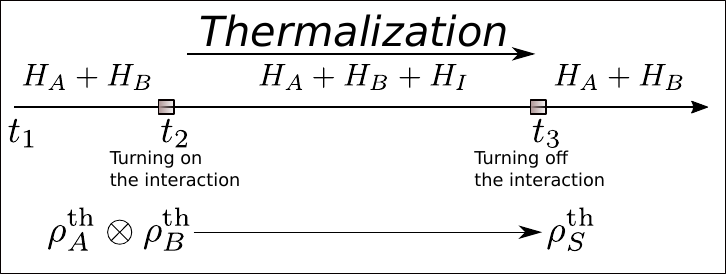}
	\caption{This figure illustrates the phases of the thermalization protocol, as described in the text.}
	\label{fig: thermalization protocol}
\end{figure}

The main idea  of our protocol is to exploit the difference between the thermal states of a bipartite system in two different cases: when the interaction between the two subsystems is turned off and when it is not.
When there is no interaction, the global thermal state is  the tensor product of the two local thermal states.
On the other hand, when the interaction is turned on, the global thermal state is more complex.
In particular, the local state of each subsystem, named \emph{reduced thermal state}, is different from the original local thermal state and has a free energy higher than the thermal one~\cite{Gallego2016,Horodecki2013} when both are computed with respect to the same free Hamiltonian of the subsystem.
This translates into formulas as follows:
\begin{equation}
\rho_A^\textup{th}\otimes\rho_B^\textup{th} \longrightarrow \rho_S^\textup{th}\neq \rho_A^\textup{th}\otimes\rho_B^\textup{th} \qq{and}
F\prt{\rho_{A\prt{B}}^\textup{rth}, H_{A\prt{B}}} > F \prt{\rho_{A\prt{B}}^\textup{th}, H_{A\prt{B}}},
\end{equation}
where we named $A$ and $B$ the two subsystems of $S$, $H_{A\prt{B}}$ is the free Hamiltonian of subsystem  $A\prt{B}$, $\rho_{A\prt{B}}^\textup{th}$ is the corresponding local thermal state and $\rho_{A\prt{B}}^\textup{rth}=\Tr_{B\prt{A}}\prtg{\rho_S^\textup{th}}$ is its reduced thermal state and $\rho_S^\textup{th}$ is computed with respect to the Hamiltonian of system $S$, $H_A + H_B + H_I$, being $H_I$  the interaction term between $A$ and $B$.
The protocol is composed of the following main steps (see {Figure}~\ref{fig: thermalization protocol} and find more information in Ref.~\cite{Piccione(2018)}).

\begin{itemize}
	\item At the start of the protocol the two subsystems of $S$ are in their local thermal states and they are non-interacting $\prt{t_1}$.
	\item Then, the resource $R$ turns on the interaction between them while not changing their states (in a small time interval preceding $t_2$, represented by the small box before $t_2$).
	\item The two subsystems of $S$, in contact with the thermal bath at temperature $T$, thermalize together to the new thermal state of $S$ (from $t_2$ to $t_3$).
	\item Lastly, the resource $R$ turns off the interaction (in a small time interval following $t_3$) while not changing the states of the subsystems of $S$.
\end{itemize}

If one follows the evolution of the free energy of the subsystems of $S$, he finds that at the end of the protocol (once the interaction is off) the extracted work, $W$,  satisfies~\cite{Piccione(2018)}
\begin{equation}
\label{eq: extracted work}
W= \Delta F\prt{\rho_A^{\textup{rth}},H_A} + \Delta F\prt{\rho_B^{\textup{rth}},H_B} + k_B T S(A:B) \leq  \ev{H_I}_{t_2} - \ev{H_I}_{t_3},
\end{equation}
where $S\prt{A:B}=S\prt{\rho_A^\textup{rth}}+S\prt{\rho_B^\textup{rth}}-S\prt{\rho_S^\textup{th}}$ is the mutual information of the $A$ and $B$ subsystems.
Moreover, the average of  $H_I$ is calculated  as soon as the interaction is on $\prt{t_2}$ and right before it is turned off $\prt{t_3}$.
We note that the term which contains the mutual information is a non-local term which, in some cases, could then be neglected because of the difficulty of its exploitation.

It is also possible to define an ideal efficiency of this process by taking into account the minimum amount of work that the system $R$ has lost.
Then, the efficiency is given by the ratio of the extracted work and the work lost by $R$:
\begin{equation}
\eta = \frac{W}{\ev{H_I}_{t_2} - \ev{H_I}_{t_3}} \leq 1.
\end{equation}

Once the thermalization protocol is completed, one can use the resulting states to perform other tasks of various kind. For example, one could want to charge an external battery.
In this case one needs a transfer protocol to move the work extracted by $S$ to the battery so that $S$ can be reinitialized and can begin another cycle of the thermalization protocol.
We studied this issue in the specific case of the application of the thermalization protocol to a system $S$ governed by the Rabi Hamiltonian at zero temperature. We will include this study in the revised version of Ref.~\cite{Piccione(2018)}.

We now apply the thermalization protocol to the Rabi model at zero temperature.

\subsection{Application to the Rabi Model at Zero Temperature}

The Rabi model describes the interaction between a two-level system and a quantum harmonic oscillator.
Its Hamiltonian has the following form~\cite{Braak2011}:
\begin{equation}
H_{\textup{Rb}}= \frac{\hbar  \omega}{2}\sigma_z+\hbar  \omega \hat{n}+\hbar  g \sigma_x \prt{a^\dagger + a},
\end{equation}
where $\sigma_z$ and $\sigma_x$ are the Pauli operators of the two-level system, $a\prt{a^\dagger}$ is the annihilation (creation) operator of the  harmonic oscillator, $\hat{n}=a^\dagger a$, $\omega$ is the frequency of both the two-level system and the harmonic oscillator (they are resonant) and $g$ is the coupling strength between the two subsystems.

In this model, the extracted work and the ideal efficiency take the following form~\cite{Piccione(2018)}:
\begin{equation}
W= \frac{\hbar\omega}{2}\prt{1+\ev{\sigma_z}_{t_3}} + \hbar \omega\ev{\hat{n}}_{t_3}=
\hbar \nu_0 + \frac{\hbar \omega}{2} - \hbar g\ev{\sigma_x \prt{a^\dagger + a}}_{t_3}
\qq{and}
\eta= -\frac{W}{\hbar g\ev{\sigma_x \prt{a^\dagger + a}}_{t_3}},
\end{equation}
where $\nu_0$ is the frequency of the ground state of the system when the interaction is on.
Both $\hbar \nu_0$ and $\hbar g\ev{\sigma_x \prt{a^\dagger + a}}_{t_3}$ are negative quantities.
Also, it is worth noting that the non-local term $S\prt{A:B}$ does not contribute to the work extraction at zero temperature.

\begin{figure}[t!]
	\centering
	\includegraphics[width=0.65\textwidth]{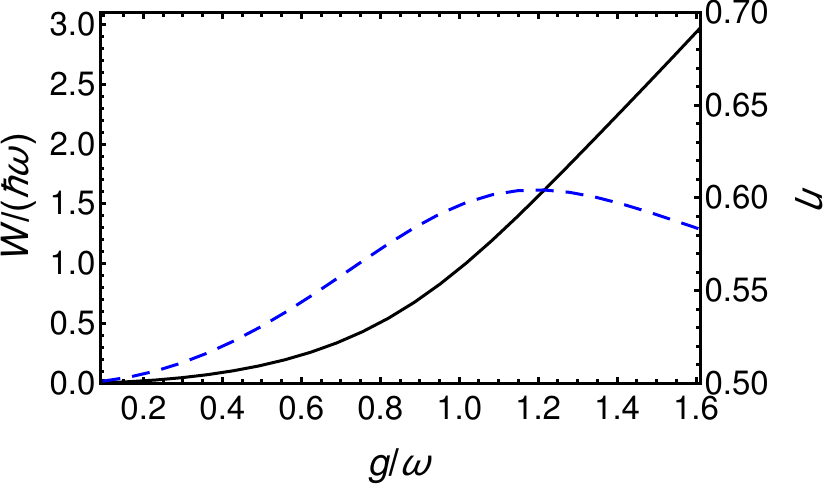}
	\caption{This graph shows the extracted work (solid black line), $W/\prt{\hbar\omega}$, and the ideal efficiency (dashed blue line), $\eta$, as a function of the coupling parameter $g/\omega$, for the Rabi model at $T=0$.}
	\label{fig:Lav-eff_d=0}
\end{figure}

The Rabi Hamiltonian has been recently exactly diagonalized~\cite{Braak2011,Chen2012}.
This let us write its ground state in the bare basis and, thus, calculate the extracted work and the efficiency of the protocol when applied to the Rabi model at zero temperature.
The graph of Figure~\ref{fig:Lav-eff_d=0} shows both the amount of extracted work and the ideal efficiency as a function of the coupling parameter in units of $\omega$.
As one can see, the extracted work is comparable with the typical energies of the two subsystems when $g/\omega$ is sufficiently high.
We remark that these high values of the coupling constant became recently experimentally attainable~\cite{Yoshihara2017}.
Moreover, the efficiency of the thermalization protocol is always higher than one half, with a peak of roughly 0.60.

These results show that the protocol we propose can achieve good results when applied to realistic physical models such as the Rabi model.

\vspace{0.2 cm}

{\bf Author Contributions:} N.P. performed the main part of the investigation and prepared a first version of the manuscript, mainly under the supervision of B.B.. All authors contributed to the conceptualization and the analysis.

\vspace{0.2 cm}

{\bf Acknowledgments:}
N.P. acknowledges the financial support of the Erasmus+ programme of the European Union and of the UTINAM Institute for the development of this programme.  B.B. acknowledges the financial support of the Observatoire des Sciences de l'Univers THETA Franche-Comt\'{e} / Bourgogne for his participation to the conference IQIS2018, including publication charges.

\vspace{0.2 cm}

{\bf Conflicts of Interest:} The authors declare no conflict of interest.


\end{document}